\documentclass[aps,pre,preprint,groupedaddress]{revtex4}

\usepackage{bm}
\usepackage{amssymb,amsmath}
\usepackage{graphicx}
\usepackage{psfrag}

\newcommand{\fullcircle}{\mbox{{\Large$\bullet$}}}

\newcommand{\fullsquare}{\mbox{$\blacksquare$}}

\begin{document}
\title{Structure and dynamics of reentrant nematics: Any open questions after
  almost 40 years?}

\author{Marco G. Mazza $^{1,\star}$ and  Martin Schoen $^{1},^{2}$}

\affiliation{%
  $^{1}$ Stranski-Laboratorium f\"ur Physikalische und Theoretische Chemie,
  Technische Universit\"at Berlin, Stra{\ss}e
  des 17. Juni 135, 10623 Berlin, Germany\\
  $^{2}$ Department of Chemical and Biomolecular Engineering, North Carolina
  State University, 911 Partners Way, Raleigh, NC 27695, U.S.A.}


\begin{abstract}
  Liquid crystals have attracted enormous interest because of the
  variety of their phases and richness of their application. The
  interplay of general physical symmetries and specific molecular
  features generates a myriad of different phenomena. A surprising
  behavior of liquid crystals is the reentrancy of phases as
  temperature, pressure, or concentration are varied.  Here, we review
  the main experimental facts and the different theoretical scenarios
  that have guided the understanding of bulk reentrant nematics.
  Recently, some computer simulations of a system confined to
  nanoscopic scales have found new dynamical features of the reentrant
  nematic phase. We discuss this prediction in relation with the
  available experimental evidence on reentrant nematics and with the
  dynamics of liquids in strongly confined environments.
\end{abstract}



\maketitle

\section{Introduction}\label{intro}

\begin{figure}
\begin{center}
\includegraphics[scale=0.4]{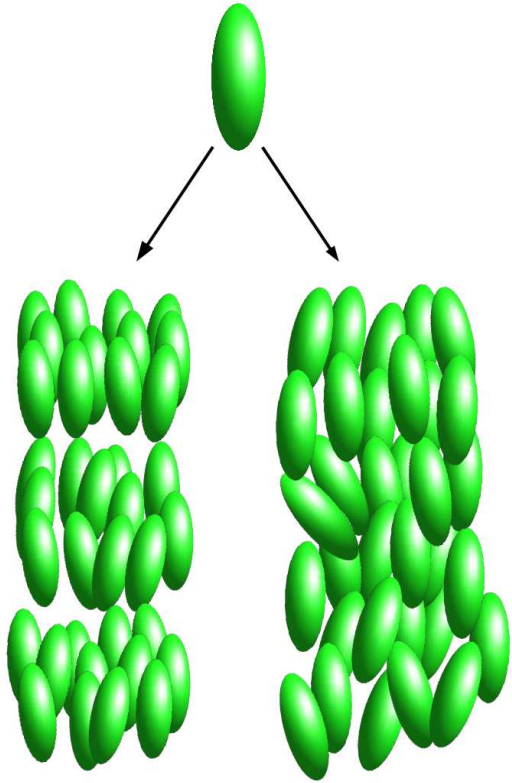}
\end{center}
\caption{Two examples of mesophases formed by rod-like molecules. Smectic A
  (bottom left) and nematic (bottom right).\label{rod}}
\end{figure}

\begin{figure}
\begin{center}
\includegraphics[scale=0.4]{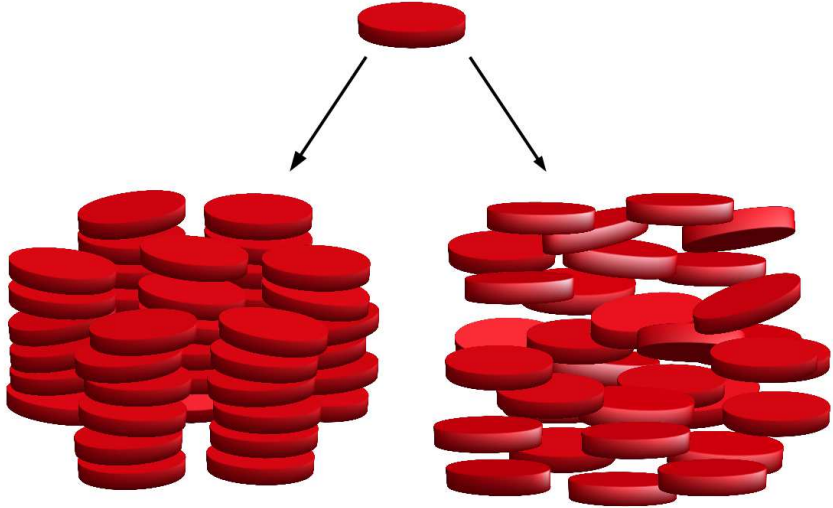}
\end{center}
\caption{Two examples of mesophases formed by disk-like molecules. Columnar
  (bottom left) and nematic discotic (bottom right).\label{disk}}
\end{figure}

In 1888 the Austrian chemist and botanist Friedrich Reinitzer observed that
the organic compound cholesteryl benzoate exhibits two melting points. He
found that it first melts at $145.5\,^{\circ}$C to a cloudy liquid, and then
at $178.5\,^{\circ}$C it becomes suddenly clear. Later, the pioneering studies
of Lehmann~\cite{lehmann}, Schenk~\cite{schenk}, Vorl\"ander~\cite{vorlaender}
and Friedel~\cite{friedel} recognized a new state of matter intermediate
between liquid and solid. Liquid crystals (LC's) were discovered. We now know
a multitude of different liquid crystalline phases, roughly divided into two
families. (i) \emph{Thermotropic} LC's undergo phase transitions as the
temperature ($T$) is changed. (ii) \emph{Lyotropic} LC's exhibit different
phases as the solvent concentration is changed.  The most commonly encountered
phases are \emph{nematic} (N), where the system has a preferential orientation
given by the nematic director $\widehat{\bm{n}}$, but the centers of mass have
no long-range correlation (see the bottom right panel of Fig.\ref{rod});
\emph{smectic} (S), where molecules organize in layers (see the bottom left
panel of Fig.\ref{rod}); \emph{columnar}, where two-dimensional order is
established (see Fig.\ref{disk}); \emph{cholesteric}, where the average
orientation spirals with a fixed pitch as different planes are traversed, and
\emph{blue phases}, which consist of three-dimensional arrangements of
cholesteric tubes.

Many different molecules form liquid crystalline phases, and their modern
importance is difficult to exaggerate. Liquid crystal displays for electronic
devices are ubiquitous. Because of the anisotropy of LC molecules and of their
interactions, LC's can easily polarize light passing through them. Highly
versatile lasers can be made with LC systems, promising numerous
applications~\cite{morris}. Liquid crystals are extremely important also for
biology. Some examples (far from being comprehensive) are the following: The
lipid bilayer of a cell membrane is a stable LC phase on which depends the
very existence of cells; strands of DNA are also known to form
LC's~\cite{rill88,clark07}; spider's silk has exceptional materials properties
due to the particular LC organization of its proteins~\cite{vollrath01};
finally, LC's are ushering a technological revolution with multiple
applications in biology and medicine as biosensors~\cite{woltman}. Many
extensive reviews of the state of the art of LC crystals exist, see for
example ~\cite{goodby} for modern applications of LC in self-assembling
supramolecular structures.

Among the interesting features of LC's that make them so versatile there is a
surprising sequence of phase transitions of rod-like molecules that still
poses challenges to both experimental and theoretical studies. The typical
sequence of LC phases encountered starts at high $T$ with an isotropic (I)
phase, where the system is characterized by complete translational and
orientational invariance.  As $T$ decreases a preferential molecular
orientation emerges -- the nematic director -- breaking the rotational
symmetry. At even lower $T$, S layers appear indicating that now
translational invariance is broken, too.

Surprisingly, in 1975 Cladis discovered \cite{cladis75} that when two
compounds, whose molecules have two benzene rings and a strongly polar cyano
group, were mixed together, a N phase formed at $T$ \emph{above} the S phase
(which is usual), but also at $T$ \emph{below} the S phase (which is
surprising). The N phase at lower $T$ was termed \emph{reentrant nematic}
(RN). The presence of a RN phase is surprising because in the vast majority of
cases as $T$ is lowered the symmetry of the stable thermodynamic phase
decreases, i.e. the state of the system is invariant under a decreasing number
of symmetry operations, eventually becoming a crystalline solid.

After the seminal work of Cladis \cite{cladis75}, other systems in a variety
of different conditions were found to have a RN phase. Today, RN behavior is a
well-established phenomenon. Here, we focus on the past and present
understanding of RN, with special attention to the dynamics of RN phases.

Naturally, nearly 40 years after Cladis' discovery of the RN phase one may
wonder: Is there still something to be understood about these phases? It is
the primary purpose of this article to demonstrate that it is indeed so.  For
exmaple, recent numerical investigations on the dynamics of LC's confined in
nanoscopic pores suggest some interesting dynamics features. In particular, it
was found \cite{mazza10,mazza11} that diffusivity in the direction of the
molecular long axis is greatly enhanced. This appears to be an analogous
situation to the levitation effect in zeolites and carbon
nanotubes~\cite{derouane,yash}. There is some favorable experimental evidence
\cite{ratna81} supporting the theoretical predictions of enhanced parallel
diffusivity~\cite{mazza10,mazza11}. However, the question is not yet settled.
Our critical review of the available experimental work on RN dynamics reveals
that numerous controversies are not yet resolved. A main problem identified
here is that reentrancy is not caused by a single microscopic mechanism, but
rather by many. Their identification is still an open question. Consequently,
the investigations of the dynamical behavior of RN's are often divergent.

The remainder of this Review is organized as follows: in
Sec.~\ref{hist} we review the seminal experiments that discovered RN phases in
both polar and nonpolar substances. In Sec.~\ref{exp} we discuss the main
experimental investigations on the nature and properties of RN's, while in
Sec.~\ref{theo} we review the different theories and scenarios put forward to
rationalize RN existence and features. In Sec.~\ref{sim} we review the more
recent computational studies of RN's, and, finally, Sec.~\ref{concl} lists our
conclusions.

\section{History}\label{hist}

The first experimental detection of RN in a LC is reported by
Cladis~\cite{cladis75}. There, an amount of hexyloxybenzilidene amino
benzonitrile (HBAB) was mixed with cyanobenzilidene butylaniline (CBOOA).
Individually, the former has a N phase for $35\,^{\circ}\mathrm{C}\lesssim
T\lesssim 101\,^{\circ}\mathrm{C}$, and the latter has a S$_\mathrm{A}$ phase
for $44\,^{\circ}\mathrm{C}\lesssim T\lesssim 83\,^{\circ}\mathrm{C}$ and a N
phase for $83\,^{\circ}\mathrm{C}\lesssim T\lesssim 108\,^{\circ}\mathrm{C}$.
When mixed together, as $T$ is lowered a sequence of I, N, S$_\mathrm{A}$, and
RN is found for a range of molar fractions $c$; the curve bounding the
S$_\mathrm{A}$ phase in the $T-c$ plane can be approximately fitted with a
parabola
\begin{equation}
c=c_0+\beta(T_\mathrm{NS}-T_0)^2,
\end{equation}
where $c_0\approx0.09$, $\beta\approx
-5.4\times10^3\,^{\circ}\mathrm{C}^{-2}$, $T_0\approx61\,^{\circ}\mathrm{C}$,
and $T_\mathrm{NS}$ is the N-S$_\mathrm{A}$ transition $T$~\cite{cladis75}.
The measurements of the bend elastic constant~\cite{cladis75} constitutes the
first examination of possible structural differences between the N and RN
phases. In this early experiment it was shown that the elastic constant has a
very similar $T$-behavior as the S$_\mathrm{A}$ phase is approached from above
or below
~\cite{cladis75}. Cladis~\cite{cladis75} emphasized that CBOOA has an
incommensurate S$_\mathrm{A}$ phase, that is, the spacing between S layers is
not an integral multiple of the molecular length. A possible way to understand
how an incommensurate spacing can arise is to imagine a S layer composed of
dimers.  Because of the strong polar character of cyano compounds molecules
naturally tend to form dimers to minimize the interaction energy; these dimers
in turn organize themselves in S layers, or, in other words, each S layer is
in reality a double layer of molecules. This initial observation is consistent
with other experiments \cite{figueirinhas} and has proven itself fruitful in
the understanding of the structure of RN phases.

Following the assumption of the double-layer nature of the S$_\mathrm{A}$
phase a series of pure compounds and some mixtures were
studied~\cite{cladis77}. Upon increasing pressure, RN phases were found in the
supercooled region of polar substances such as 4-cyano-4'-octyloxy biphenyl
(COOB) or CBOOA. What these substances have in common is the amphiphilic
nature of their molecules. Unlike most LC's the polar part (the aromatic
rings) is not in the middle of the chains, but rather at one end. The polar
cyano group has a dipole moment of $4.5$~D. The nonpolar part (aliphatic
chain) forms the opposite end of the amphiphile. This structure will naturally
favor a certain degree of dimerization, since the polar groups will
preferentially interact with each other through long-range forces, and the
nonpolar tails through short-range forces.  Cladis \emph{et
  al.}~\cite{cladis77} proposed that it is the short-range interaction between
nonpolar tails what stabilizes the S layers. This hypothesis is successfully
tested by measuring the dependence of the maximum stability pressure $P_m$ of
the S$_\mathrm{A}$ on the number of carbon atoms in the nonpolar tails. It is
found that $P_m$ increases linearly as the number of methylene groups
increases. It is argued~\cite{cladis77,cladis77pra} that as pressure increases
(or equivalently as $T$ decreases) the interaction between the polar groups
becomes repulsive, and at the same time the nonpolar tails are somewhat
compressed, lowering the energy barrier to permeation through the S layers.
Both effects lead to a destabilization of the S order, and hence to the
formation of a RN.

After 1979, the occurrence of RN could not be dismissed as a peculiarity of
few molecules, since a stable RN at atmospheric pressure was found for
molecules exhibiting three benzene rings, and, importantly, not in mixtures
but in pure compounds~\cite{hardouin79,sigaud79,sigaud81}.  Additionally,
reentrant \emph{smectic} phases were also found ~\cite{hardouin79ssc,tinh83}.
In the following years different groups found many different systems
exhibiting RN phases (see, e.g., the work of Sigaud \emph{et
  al.}~\cite{sigaud81} and references therein), and also multiple reentrant
phases~\cite{tinh82,tinh84}.  It has been speculated that quadruple reentrance
is possible if compounds with four or five benzene rings are
employed~\cite{indekeu86}.  Cladis's own account of the discovery and
thermodynamic understanding of RN phases provides many more
details~\cite{cladis88}.

Until 1983 it was believed that only molecules with a strong polar group
showed a reentrant behavior. However, Diele \emph{et al.} showed the existence
of RN phases in terminal-nonpolar substances~\cite{diele83,pelzl}.
Importantly, their X-ray measurements showed that the thickness of the S
layers is equal to the average molecular length. Hence, dimerization cannot
take place in these nonpolar compounds and the occurrence of a RN phase must
have a different physical origin~\cite{diele83,pelzl} (see Sec.~\ref{theo} for
the theories applicable to nonpolar systems).

A question naturally arising is whether RN and N phases are different. One can
consider two aspects: structure and dynamics. In the following we explore the
experimental evidence addressing this question. We can, however, anticipate
that, since pure compounds and mixtures, polar and nonpolar molecules all
exhibit RN phases, the picture emerging will be rather complex and diverse.

\section{Experimental investigations}\label{exp}

Structural differences of the N and RN phases were investigated with X-ray
scattering~\cite{guillon78} for mixtures of
N-(\emph{p}-hexyloxybenzylidene)-\emph{p}-aminobenzonitrile (HBAB) and
N-\emph{p}-cyanobenzylidene-\emph{p}-\emph{n}-octyloxyaniline. No qualitative
difference was found between the N and RN phases, i.e. the Bragg peak
indicating the layer thickness appears always in the range $33-35$\AA;
however, for some mixtures the RN was found to coexist with microcrystallites.
Guillon \emph{et al.} concluded that these crystalline fluctuations share the
same structural properties and are energetically similar to the RN
phase~\cite{guillon78}.

Nuclear magnetic resonance (NMR) offers a number of techniques to investigate
both structure and dynamics. The first NMR study of a pure LC system with a RN
phase was carried out by Miyajima \emph{et al.}~\cite{miyajima84}.  For pure
4-($4''$-octyloxybenzoloxy)-benzylidene-$4'$-cyanoaniline (OBBC) they found a
smooth continuous increase in the orientational order parameter across the N,
S$_\mathrm{A}$, and RN phases. However, when studying a binary mixture with a
spin-probe impurity probe they found a steep increase of the nematic order
parameter in the RN phase. While proton NMR is sensitive to the aromatic core,
the impurity probe method is not. Therefore, the conflicting results in the
orientational order parameter can be ascribed to the different techniques.
From the $T$ behavior of the spin-lattice relaxation time $T_1$ measured at
$29.8$~MHz for OBBC Miyajima \emph{et al.} conclude that translational
self-diffusion plays a dominant role in $T_1$ relaxation in the RN phase (with
a characteristic Arrhenius $T$-dependence), while the N phase $T_1$ behavior
is dominated by nematic director fluctuations.

However, for a LC mixture Dong reached the opposite conclusion that self-diffusion appears to
be the relevant relaxation mechanism in the N phase, while nematic director
fluctuations dominate the RN phase \cite{dong81a,dong81,dong82}.  Using proton
NMR he studied the $T$ dependence of $T_1$ for a LC binary
mixture~\cite{dong81a,dong81} and found an Arrhenius behavior in the N phase,
and no $T$-dependence in the RN phase within experimental error; that director
fluctuations dominate the relaxation of the RN phase is further corroborated
by the linear relationship between $T_1^{-1}$ and $\omega^{-1/2}$ where
$\omega$ is the angular frequency, as predicted by the theory of spin
relaxation through a  director-fluctuation mechanism \cite{dong82}.

We cannot avoid stressing at this point something that has not been clearly
stated in previous works. Although reentrancy is found in many different
substances, their dynamics need not be the same. Specifically, pure compounds
and mixtures do not necessarily have to exhibit the same dynamical properties.
Even though there are discrepancies in the experimental results, we
can safely conclude that the dynamics of N and RN are markedly different.

A somewhat intermediate situation is described by Sebasti\~ao \emph{et
  al.}~\cite{sebastiao} who performed proton NMR experiments on a polar LC
compound which exhibits N, partial bilayer S, RN, and reentrant S$_\mathrm{A}$
phases.  Self-diffusion is shown to contribute little to $T_1$ in the N and RN
phase. Instead, fluctuations of the director and reorientations are the main
dynamical processes contributing to $T_1$.  Interestingly, however, the ratio
between the translational self-diffusion parallel and perpendicular to the
director is slightly higher in the RN phase than in the N phase
~\cite{sebastiao}. Those results~\cite{sebastiao} are consistent with a
dynamical process of association and dissociation of molecular groups, which
are important in the double-layer S and N phases.

By using ${}^{129}$Xe NMR Bharatan and Bowers \cite{bharatam99} studied two LC
mixtures and found that $T_1$ (and also the spin-spin relaxation time $T_2$)
shows Arrhenius behavior in both N and RN phases. The activation energies
extracted from the data do not behave in a similar fashion for the two systems
studied. For a binary mixture, the activation energy in the RN is more than
two times the value in the N phase, whereas for a ternary mixture, there is
almost no difference in the activation energies.  The increase in activation
energy for the first LC mixture is ascribed to changes in the molecular
packing in the RN phase~\cite{bharatam99}.

Although a number of other studies were carried out on mixtures and pure
compounds with a RN phase~\cite{dong87,dong98,dong00} using deuteron NMR
little can be said about translational self-diffusion, because this technique
is sensitive to other types of motion, e.g. intramolecular rotations,
reorientational dynamics, or supramolecular motion, such as order director
fluctuations. The picture arising from the work of Dong and collaborators is
one of very subtle changes between the phases, and especially between N and RN
phases.

The structure of RN's also poses some unanswered questions. For example, is
there a change in nematic order parameter $Q$ at the transition from S to RN?
The work of Vaz \emph{et al.}~\cite{vaz83} shows that $Q$ increases upon
entering the RN. They measured the $T$-dependence of $Q$ in a LC
binary mixture with deuterium NMR and found that at the S$_\mathrm{A}$-RN
transition $Q$ displays a sharp increase in slope~\cite{vaz83}.  To
rationalize this behavior they considered an expansion of the free energy
difference $\Delta F\equiv F_N+F_{S}-F_I$ between the N, S and I phase
in terms of $Q$ and a smectic order parameter $\Psi$, that to leading
order is
\begin{equation}
\Delta F=\frac{1}{2}\alpha Q^2+\frac{1}{3}\beta Q^3+\frac{1}{4}\gamma Q^4 +
\frac{1}{2}a\Psi^2+\frac{1}{4}b\Psi^4,
\end{equation}
where $\alpha=\alpha_0(T-T_{IN})$, $a=a_0(T-T_{NA})$, $T_{IN}$ and $T_{NA}$
are the I-N and the N to S$_\mathrm{A}$ transition $T$, respectively, and all
the constants are positive except $\beta<0$. This expansion of $\Delta F$
produces a first order I to N phase transition, and a second order N to
S$_\mathrm{A}$ transition.  Following Cladis~\cite{cladis81}, they assumed a
quadratic coupling of the order parameters proportional to $Q^2\Psi^2$ to
describe the RN phase and derived that upon entering the RN from the
S$_\mathrm{A}$ phase the nematic order parameter should be positively
perturbed ($\delta Q>0$) such that the phenomenological description is in
agreement with the experiments~\cite{vaz83}.

However, using deuterium NMR on a binary mixture, Emsley \emph{et
  al.}~\cite{emsley} found only subtle changes in the order parameters at the
S$_\mathrm{A}$-RN phase transition. Contrasting results are also presented by
Dong \emph{et al.}~\cite{dong85}, where deuterium NMR experiments
showed no enhanced orientational order in the RN phase, in contradiction with
both a Landau free energy expansion theory and a McMillan theory of RN's.

Electron spin resonance experiments on the LC mixture of hexyloxy biphenyl
(6OCB) and octyloxy biphenyl (8OCB) found no dramatic structural changes at
the S$_\mathrm{A}$-RN transition using a number of different spin probe
molecules and similar dynamics of the probes in reentrant and nonreentrant
LC~\cite{nayeem89}. Nayeem and Freed also concluded, contrary to Dong \emph{et
  al.}~\cite{dong87}, that their results are ``\emph{not consistent with
  saturation in pair formation being a necessary precursor to
  reentrance}''~\cite{nayeem89}.  The differences in these experimental
results can be ascribed to different experimental techniques sensitive to a
different extent to the orientational degrees of freedom of the aromatic cores
or of the molecular chains. Nonetheless, the question still lingers: What is
the behavior of the nematic order across the S-RN phase transition?

A very different experimental approach was adopted by Ratna \emph{et
  al.}~\cite{ratna81} who measured the electrical conductivity in the
direction parallel ($\sigma_\parallel$) and perpendicular ($\sigma_\perp$) to
the optic axis, which in the orientationally ordered phases coincides with the
nematic director. They studied the pure compound
4-cyanophenyl-$3'$-methyl-4($4'$-n-dodecylbenzoyloxy)benzoate (12 CPMBB) and
the mixture 6OCB/8OCB. For the single component system they found a remarkable
increase of the ratio $\sigma_\parallel/\sigma_\perp$ as the system
approaches the RN-S$_\mathrm{A}$ transition temperature $T_{RN-A}$. At the
lowest $T$ studied for the pure compound ($1.2\,{}^\circ$C below $T_{RN-A}$)
$\sigma_\parallel$ is approximately equal to $16\sigma_\perp$. For the mixture
6OCB/8OCB, they found the same qualitative increase of the ratio 
$\sigma_\parallel/\sigma_\perp$, but at the lowest $T$ studied in the RN
phase the value of this ratio was only $1.8$~\cite{ratna81}. Although the
results for the mixture would require additional analysis, what was found for
the pure compound is strong evidence supporting a scenario of increased mass
transport in the RN phase.

Another insight into the differences between N and RN phases comes from the
work of Nozaki \emph{et al.}~\cite{nozaki} who found that for the mixture
6OCB/8OCB the rotations around the short molecular axis are greatly hindered
in the RN phase with respect to the N phase. Specifically, by measuring the
$T$-dependence of the complex permittivity with time-domain reflectometry,
they found a higher value of the activation energy in the RN compared with the
N phase.


There are limited modern investigations of the RN state. Yethiraj \emph{et
  al.} measured proton NMR spectra of solutes dispersed in a mixture of
6OCB/8OCB to determine the smectic and nematic order parameters in the
Kobayashi-McMillan theory~\cite{yethiraj}. Das and Prasad \cite{das2011}
measured the rotational viscosity of a LC binary mixture and found similar
values in the N and RN phases. However, there are strongly contrasting results
\cite{bhattacharya80,dong81} suggesting that the rotational viscosity of RN
may be two or three orders of magnitude higher than in the N phase.

Finally, hydrodynamic flow in RN's has received some attention. Bhattacharya
and Letcher measured capillary shear flow for a ternary mixture with a RN
phase. They found that the N and RN phases have identical flow properties, and
the hydrodynamic theory of N phases can be applied to RN
phases~\cite{bhattacharya80}.  Also, Jad\.zyn and
Czechowski~\cite{jadzyn} measured the viscosity for the well studied mixture
6OCB and 8OCB, and found a distinct decrease of viscosity upon entering the RN
phase.

\section{Theoretical scenarios for reentrancy}\label{theo}

The experimental findings discussed in Sec.~\ref{exp} well illustrate the
complexity of RN phases. It is therefore not surprising that a similar rich
picture also emerges from theoretical studies. Prost and Barois derived a
Landau theory of LC exhibiting reentrancy from general considerations of
symmetry and on the basis of the relevant length scales involved
\cite{prost83,prost84}. In a S phase a density modulation appears. On average
the molecular centers lie on parallel planes. Thus, it is natural to consider
the expansion of the density in a Fourier series
\begin{equation}
\rho=\rho_0+\frac{1}{\sqrt{2}}(\psi e^{i\mathbf{q}\cdot\mathbf{r}})+\ldots
\end{equation}
where the wave-vector $\mathbf{q}=(2\pi/l)\widehat{\bm{n}}$, $l$ is the layer
thickness, and $\widehat{\bm{n}}$ is the director. However, polar substances
and mixtures exhibiting RN phases show two incommensurate periodicities in
X-ray scattering experiments, i.e. one periodicity equal to the molecular
length $l$, and a second one $l'$ that is not an integer multiple of the
first, typically $l'\approx1.2$ -- $1.5\,l$.  It is necessary to introduce the
total dipole moment $\mathbf{P(r)}$ to describe a system of polar molecules
\begin{equation}
\mathbf{P(r)}\equiv\frac{1}{V}\sum_i\mathbf{p}_i\delta(\mathbf{r}-\mathbf{r}_i)\,,
\end{equation}
where $\mathbf{p}_i$ is the dipole moment of molecule $i$. A potential can be
derived from $\mathbf{P(r)}$ that may be used as a second order
parameter
\begin{equation}
\mathbf{P(r)}=\frac{1}{4\pi}\nabla\phi\,.
\end{equation}
With these two order parameters, $\rho$ and $\phi$, a Landau-Ginzburg free
energy functional can be written as
\begin{equation}
F=\frac{1}{2}\int \mathrm{d}^3\mathbf{r} \{ A_\rho\rho^2+A_\phi\phi^2-D\rho\phi^2+
\frac{1}{2}B_\rho\rho^4+\frac{1}{2}B_\phi\phi^4+C\rho^2\phi^2 \} \,.
\label{GL}
\end{equation}
From the competition between the different length scales contained in
Eq.~(\ref{GL}), Prost and collaborators \cite{prost83,prost84} obtained a
phase diagram with different S and also RN phases.

The first microscopic theory of the physical origin of RN phases was
elaborated by Berker and Walker in 1981~\cite{berker81}. They started from the
observation that most reentrant nematogens are composed of molecules with a
strong polar group at one end and an attached aliphatic chain forming the
other end of the molecule. The dipolar interaction is assumed to be
antiferroelectric.  Considering a plane normal to the nematic director,
antiferroelectric long-range order cannot be established in a periodic regular
arrangement of molecules. However, Berker and Walker argued that in a liquid
molecules can relax their positional degrees of freedom and configurations of
triplets with two short antiferroelectric bonds and one unfavorable
ferroelectric, but longer bond will be possible. In this way frustration is
avoided and the system can condense to a liquid of dimers forming a
two-dimensional network.  Hence, a bilayer S phase forms. As $T$ is lowered
positional order arises that leads to increasing frustration and to the
breaking of the dimers. As a consequence a nematic phase appears again. The
interaction of two dipoles $\mathbf{s}_1$ and $\mathbf{s}_2$ separated by a
distance $|\mathbf{r}_{12}|$ is described by the potential
\begin{equation}
U(\mathbf{r}_1,\mathbf{s}_1,\mathbf{r}_2,\mathbf{s}_2)=\frac{A \mathbf{s}_1\cdot \mathbf{s}_2- 3B(\mathbf{s}_1\cdot \mathbf{r}_{12})(\mathbf{s}_2\cdot \mathbf{r}_{12})}{|\mathbf{r}_{12}|^3}\,,
\end{equation}
where $|\mathbf{s}_i|=1$. For $B<A$ the interaction is antiferroelectric,
whereas for $B=A$ it is purely dipolar, and for $B>A$ the interaction becomes
ferroelectric. By integrating out the positional degrees of freedom and
distinguishing different degrees of interaction, Berker and Walker mapped the
initial system onto an Ising model with annealed bond disorder. The phase
diagram obtained~\cite{berker81} shows a RN phase for a range of model
parameters and also a doubly reentrant phase sequence. Berker \emph{et
  al.}~\cite{indekeu-pra86,indekeu86} extended their theory to describe
multiple reentrant behavior, considering partial bilayer (S$_\mathrm{Ad}$) and
monolayer (S$_{\mathrm{A}1}$) smectic phases, i.e.
N--S$_\mathrm{Ad}$--N--S$_\mathrm{Ad}$--N--S$_{\mathrm{A}1}$, which was found
experimentally in at least one pure substance and its
mixture~\cite{tinh-jphysfr82} and S$_\mathrm{C}$ phases~\cite{netz92}. The
scenario of frustrated dipolar interaction~\cite{indekeu-pra86,indekeu86}
(commonly referred to as ``spin-gas theory'') has played a major role in
understanding different instances of reentrancy in LC~\cite{cladis88}. An
implication of the scenario described above is that the N-S$_\mathrm{A}$
transition is different from the S$_\mathrm{A}$-RN transition. This is at odds
with some experiments that show very subtle or no structural difference
between the nematic phases for LC mixtures~\cite{nayeem89,kortan81} and a pure
compound~\cite{guillon80}.

A different approach to a microscopic theory for reentrancy in nematogens was
proposed by Longa and de Jeu~\cite{longa82}: Pairs of molecules with
antiparallel dipoles associate to form dimers in dynamical equilibrium with
monomers. As $T$ is decreased the fraction of dimers in the LC system
increases and dipolar dispersive forces between the dimers lead to the
formation of a S$_\mathrm{A}$ phase. It is important to note that monomers
under these conditions help stabilizing the S$_\mathrm{A}$ phase because
monomers can fill the voids left by the bulkier dimers. However, as the dimer
fraction increases it becomes increasingly difficult to pack them without
perturbing the S layers.  It is instead, as Longa and de Jeu
argue~\cite{longa82}, entropically more favorable to disrupt the S layering by
forming a RN phase.  By extending McMillan's theory of the N--S$_\mathrm{A}$
transition~\cite{mcmillan}, Longa and de Jeu used a mean-field approach to
solve for the one-particle distribution functions of monomers and dimers.
Although Berker and
collaborators~\cite{berker81,indekeu-pra86,indekeu86,netz92} too considered
dipolar interactions to be fundamental, they viewed the S phase as a
percolating network of antiparallel pairs where frustration is avoided only
because of liquid positional disorder. In our view, contrary to the
conclusions of Figueirinhas \emph{et al.}~\cite{figueirinhas}, it is difficult
to exclude either theory in favor of the other, because the fundamental
experimental predictions of the spin-gas theory, and of the McMillan-like
theory are quite similar: They both predict a roughly $T$-independent value
for the smectic layer thickness and antiparallel associated pairs.

It is worth mentioning other approaches to explain reentrancy in LC. For
example, Luckhurst and Timini~\cite{luckhurst81} found a N $\to$
S$_\mathrm{A}$ $\to$ RN phase sequence upon decreasing $T$. They only assumed
a linear $T$ dependence of the parameter $\alpha$ that in the original
McMillan theory~\cite{mcmillan} is empirically related to the length of the
molecular alkyl chain. In practice, as $\alpha$ decreases with decreasing $T$,
the particles described by the theory turn from dimers to monomers.  Ferrarini
\emph{et al.}  \cite{ferrarini96} showed that reentrant phases can also be
recovered by considering chemical reactions of isomerization and dimer
formation under reasonable conditions.

All the models and theories discussed above refer to polar molecules. It it
natural that most theoretical work has focused on this class of LC's, because
experimentally reentrancy is predominantly found in LC molecules with strong
dipolar groups (typically cyano). However, RN phases have also been observed
in nonpolar substances \cite{diele83,pelzl}. Dowell~\cite{dowell} devised a
lattice theory for a single component system that focuses on molecular chain
flexibility. In Dowell's model the S phase is formed because of segregated
packing of cores and tails. As $T$ is lowered, however, the molecular chains
become increasingly more stiff, destabilizing the S layers. Eventually, it
becomes entropically more favorable to disrupt S layering such that a RN phase
is formed. X-ray \cite{kortan81} and ESR \cite{nayeem89} experiments support
Dowell's scenario of reentrancy in systems that do not show signs of
dimerization~\cite{diele83,pelzl}.  Bose \emph{et al.}  \cite{bose} extended
McMillan's theory to include chain flexibility and also obtained a RN for a
single component nonpolar system.

\section{Computer Simulations}\label{sim}

In modern times vast possibilities of theoretical investigation have been
ushered by the growth of computational power and the sophistication of
simulation algorithms.  More and more complex fluids and microscopic
structures can now be simulated~\cite{gubbins10}. Computer simulations were
first applied to the study of RN's by Netz and Berker~\cite{netz92}. They
solved numerically the lattice spin-gas model introduced in Sec.~\ref{theo}
with Monte Carlo (MC) calculations. The first off-lattice MC simulations were
performed only in 2005 by de Miguel and Mart\'in del R\'io~\cite{demiguel05}.
They considered a single-site molecular model for mesogens, where molecules
are represented by parallel hard-core ellipsoids with a spherically symmetric
square well potential (around each molecular long axis).  In that study a RN
phase and two tricritical points are found, one located on the
N-S$_\mathrm{A}$ transition line and the second one on the S$_\mathrm{A}$-RN
transition line.  In this simple model, the presence of a RN in a nonpolar LC
is rationalized in terms of entropy. In the S$_\mathrm{A}$ phase the energy is
minimized and within the fluid state no further decrease is possible.
However, the free energy can still be minimized by an increase in entropy.
Thus, the loss of positional order is entropically driven~\cite{demiguel05}.

Though the system of parallel hard-core ellipsoids~\cite{demiguel05} offers a
clear picture of the origin of reentrancy in nematics, it has the shortcoming
of neglecting rotational dynamics altogether, because the molecular long axes
point in the same fixed direction. Suppressing orientational dynamics leaves
open the question of whether the RN in such simple models is robust against
orientational fluctuations.


\subsection{Confined Mesogens }

Confined fluids have become an extremely active field of research in recent
years \cite{schoen07,schoen10,schoenpccp08}.  Novel phenomena take place in
liquids confined to narrow spaces, e.g. wetting, layering and many others
induced by the interaction with the confining surface. The interplay of these
new phenomena with the physics of bulk fluids gives the possibility of both
testing our understanding of molecular mechanisms, and devise concrete
applications not otherwise feasible. In fact, applications of confined fluids
range from nanotechnology to biomedical devices.  It is therefore natural that
LC properties and phase transitions come under a new scrutiny. Furthermore,
nanoconfinement is a physically sensible (and realistic) way of inducing
ordering fields into the fluid system.

Below, we review computer simulations of a pure LC system confined between
walls separated by a nanoscopic distance (few tens of molecular diameters)
performed by the present authors and their colleagues~\cite{mazza10,mazza11}.
These are the first simulations to fully take into account a three dimensional
fluid with rotational degrees of freedom.

\subsubsection{Model}
Because the focus is on the dynamical behavior of the LC's, molecular dynamics
(MD) was employed.  Specifically, we \cite{mazza10,mazza11} adopted the
Gay-Berne-Kihara (GBK) model for prolate mesogens \cite{martinez04}. The GBK
model potential conveniently takes into account both the anisotropy of mesogen
interaction and the spherocylindrical molecular shape \cite{note1} which is
considered to approximate prolate mesogens better than ellipsoids
\cite{martinez04}.

In the GBK model the interaction between a pair of spherocylinders $i$ and $j$
depends on the molecular orientations represented by the unit vectors
$\widehat{\bm{u}}_i$ and $\widehat{\bm{u}}_j$, respectively and their distance
$\bm{r}_{ij}\equiv\bm{r}_{i}-\bm{r}_{j}$. Specifically
\begin{equation}\label{eq:ljmod}
u_{\mathrm{ff}}=
4\varepsilon_{\mathrm{ff}}(\widehat{\bm{r}}_{ij},\widehat{\bm{u}}_{i},\widehat{\bm{u}}_{j})
\left[
\left(\frac{\sigma}{d_{ij}^{\mathrm{m}}}\right)^{12}-
\left(\frac{\sigma}{d_{ij}^{\mathrm{m}}}\right)^{6}
\right]
\end{equation}
where $\widehat{\bm{r}}\equiv\bm{r}/r$, $r\equiv\left|\bm{r}\right|$, and the
function
$d_{ij}^{\mathrm{m}}(\bm{r}_{ij},\widehat{\bm{u}}_{i},\widehat{\bm{u}}_{j})$
is the {\em minimum} distance between that pair of molecules~\cite{vega94}.
The orientation-dependent interaction strength in Eq.~(\ref{eq:ljmod}) is
given by
\begin{eqnarray}\label{eq:epsilongb}
\varepsilon_{\mathrm{ff}}(\widehat{\bm{r}}_{ij},\widehat{\bm{u}}_{i},\widehat{\bm{u}}_{j})&=&
\epsilon_{\mathrm{ff}}
\left\{
1-
\frac{\chi^{\prime}}{2}
\left[
\frac{(\widehat{\bm{r}}_{ij}\cdot\widehat{\bm{u}}_{i}+\widehat{\bm{r}}_{ij}\cdot\widehat{\bm{u}}_{j})^2}{1+\chi^{\prime}\widehat{\bm{u}}_{i}\cdot\widehat{\bm{u}}_{j}}+
\frac{(\widehat{\bm{r}}_{ij}\cdot\widehat{\bm{u}}_{i}-\widehat{\bm{r}}_{ij}\cdot\widehat{\bm{u}}_{j})^2}{1-\chi^{\prime}\widehat{\bm{u}}_{i}\cdot\widehat{\bm{u}}_{j}}\right]\right\}^2\nonumber\\
&&\times
\frac{1}{\sqrt{1-\chi^2(\widehat{\bm{u}}_{i}\cdot\widehat{\bm{u}}_{j})^2}}\,, 
\end{eqnarray}
where the parameters $\chi$ and $\chi^{\prime}$ are given by
\begin{subequations}\label{eq:chichip}
\begin{eqnarray}
\chi&\equiv&
\frac{\kappa^2-1}{\kappa^2+1}\label{eq:chi}\\
\chi^{\prime}&\equiv&
\frac{\sqrt{\kappa^{\prime}}-1}{\sqrt{\kappa^{\prime}}+1}\,.\label{eq:chip}
\end{eqnarray}
\end{subequations}
In these last two expressions parameters $\kappa=L+1$ ($L$ in units of
$\sigma$), and $\kappa^{\prime}$ represents the interaction strength for a
side-side relative to an end-end configuration of a pair of spherocylinders.

We consider a LC system confined along the $z$ direction by two planar,
structureless walls. We model the fluid-substrate interaction via
\begin{equation}\label{eq:ufs}
u_{\mathrm{fs}}=
4\epsilon_{\mathrm{fs}}\rho_{\mathrm{s}}
\left[
\left(\frac{\sigma}{d_{ik}^{\mathrm{m}}}\right)^{10}-
\left(\frac{\sigma}{d_{ik}^{\mathrm{m}}}\right)^{4}
g(\widehat{\bm{u}}_i)
\right]
\end{equation}
where the parameter $\epsilon_{\mathrm{fs}}$ is the strength of
fluid-substrate interaction and $\rho_{\mathrm{s}}=2/\ell^2$ is the areal
density of the substrate where $\ell/\sigma=1/\sqrt[3]{4}$ is the lattice
constant of a single layer of atoms arranged according to the ($100$) plane of
the face-centered cubic lattice~\cite{mazza10}.  The diameter $\sigma$ of
these substrate atoms is taken to be the same as the diameter of a
spherocylinder of the confined fluid phase.

As already mentioned, a solid substrate is also a physically realistic way of
inducing an ordering field in a LC system. A number of experimental techniques
were devised to realize specific substrate features, and especially anchoring.
Some example are rubbing, polishing, exposing to photolithographic relief, or
evaporation of oxide films \cite{sonin}.  In Eq.~(\ref{eq:ufs}), $0\le
g(\widehat{\bm{u}})\le1$ is the so-called ``anchoring function''.  It allows
one to discriminate energetically different orientations of a molecule with
respect to the substrate plane. We employed degenerate \emph{planar}
anchoring~\cite{jerome}
\begin{equation}
g(\widehat{\bm{u}})=
(\widehat{\bm{u}}\cdot\widehat{\bm{e}}_{\mathrm{x}})^2+
(\widehat{\bm{u}}\cdot\widehat{\bm{e}}_{\mathrm{y}})^2
\end{equation}
where $\widehat{\bm{e}}_{\alpha}$ ($\alpha=x, y, z$) is a unit vector pointing
along the $\alpha$-axis of a space-fixed Cartesian coordinate system. Hence,
any molecular arrangement parallel with the substrate plane is energetically
favored, whereas a homeotropic alignment of a molecule
($\widehat{\bm{u}}\parallel\widehat{\bm{e}}_{\mathrm{z}}$) receives an energy
penalty by ``switching off'' the fluid-substrate attraction altogether.

The MD simulations were carried out using a velocity-Verlet algorithm for
linear molecules \cite{ilnytskyi02}. After equilibrating the system to the
chosen pressure and $T$, the fluid system is simulated in the microcanonical
ensemble to obtain information on its dynamics~\cite{mazza11}. We simulated
systems containing up to $N=1500$ molecules.  Quantities of interest will be
expressed in the customary dimensionless (i.e., ``reduced'') units. For
example, length will be expressed in units of $\sigma$, energy in units of
$\epsilon_{\mathrm{ff}}$, temperature in units of
$\epsilon_{\mathrm{ff}}/k_{\mathrm{B}}$, time in units of
$(\sigma^2m/\epsilon_{\mathrm{ff}})^{1/2}$ using $m=1$, and pressure
$P_{\parallel}$ in units of $k_{\mathrm{B}}T/\sigma^3$ where
$P_{\parallel}=\frac{1}{2}(P_{\mathrm{xx}}+P_{\mathrm{yy}})$ is related to
diagonal components of the pressure tensor $\mathbf{P}$ acting in the $x$--$y$
plane.

\subsubsection{Results}

\begin{figure}
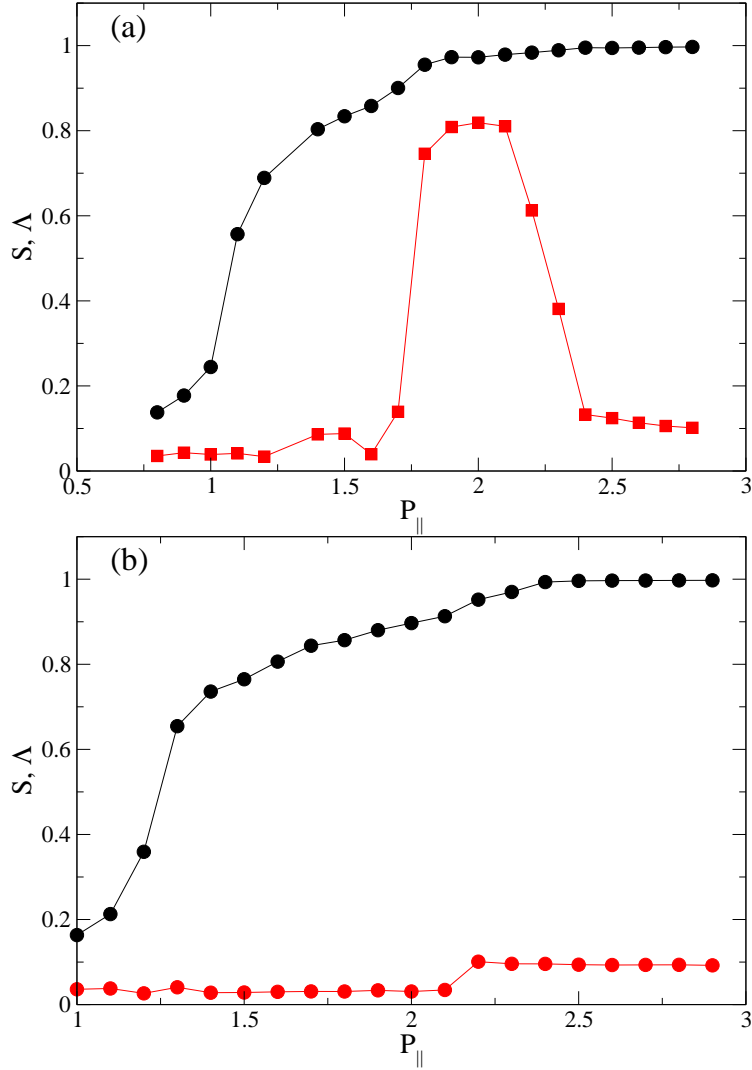

\begin{center}
\includegraphics[width=0.6\linewidth]{nem-sm-CompAFxy-T4.0.eps}
\includegraphics[width=0.6\linewidth]{nem-sm-CompAFxy-T6.0.eps}
\end{center}
\caption{Color online) Plots of nematic $S$ ($\fullcircle$) and smectic order
  parameter $\Lambda$ ($\fullsquare$) as functions of pressure $P_{\parallel}$
  at (a) $T=4.0$, and (b) at $T=6.0$.}\label{fig4}
\end{figure}

To characterize the structure of N, S$_\mathrm{A}$, and RN phases a suitable
order parameter is the so-called alignment tensor defined as
\cite{degennes71,pardowitz80}
\begin{equation}\label{eq:align}
\mathbf{Q}\equiv
\frac{1}{2N}
\sum\limits_{i=1}^{N}
\left(
3\widehat{\bm{u}}_{i}\otimes\widehat{\bm{u}}_{i}-\mathbf{1}
\right)
\end{equation}
where ``$\otimes$'' denotes the direct (i.e., dyadic) product and $\mathbf{1}$
is the unit tensor. Hence, $\mathbf{Q}$ is a real, symmetric, and traceless
second-rank tensor which can be represented by a $3\times3$ matrix. Its
largest eigenvalue is commonly used as the N order parameter $S$.

To characterize the layering of the fluid in the direction of
$\widehat{\bm{n}}$ characteristic of S phases a suitable order parameter is
given by the leading coefficient of the Fourier series of the local density
$\rho(\bm{r})$
\begin{equation}\label{eq:Lambda}
\Lambda\equiv
\frac{1}{N}
\left\langle
\left|
\sum\limits_{i=1}^{N}
\exp\left[
\frac{2\pi i\left(\bm{r}_{i}\cdot\widehat{\bm{n}}\right)}{d}
\right]
\right|
\right\rangle
\end{equation}
where $d$ is the spacing between adjacent S$_\mathrm{A}$ layers. From
Eq.~(\ref{eq:Lambda}) it is also apparent that $\Lambda\in\left[0,1\right]$,
where $\Lambda\approx0$, and $\Lambda\approx1$ in  ideal N and
S$_\mathrm{A}$ phases, respectively.

Plots of the dependence of $S$ and $\Lambda$ on $P_{\parallel}$ are presented
in Fig.~\ref{fig4} for $T=4.0$ and $6.0$.  Both $S$ and $\Lambda$ are small at
low pressures characteristic of the I phase.  Here we adopt a value
$S\simeq0.4$ as a heuristic threshold for the N phase~\cite{maier59,maier60}.
From Fig.~\ref{fig4} it therefore appears that the N phase forms somewhere
above $P_{\parallel}\simeq1.0$ ($T=4.0$) and $1.2$ ($T=6.0$), respectively.

\begin{figure}
\begin{center}
\includegraphics[width=0.6\linewidth]{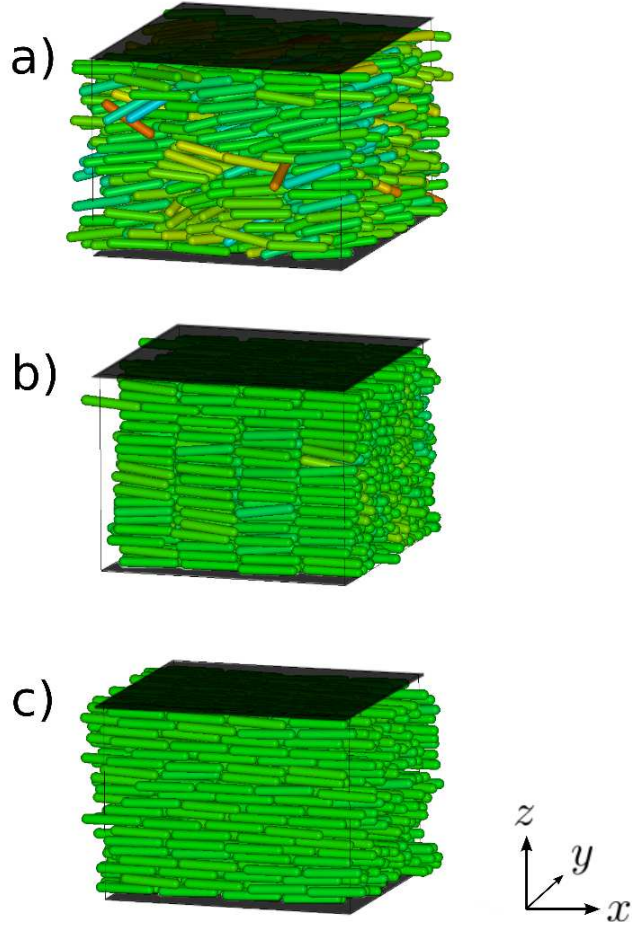}
\end{center}
\caption{(Color online) ``Snapshots'' of characteristic configurations at
  $T=4.0$; (a) $P_{\parallel}=1.6$ (N), (b) $P_{\parallel}=2.1$
  (S$_\mathrm{A}$ ), (c) $P_{\parallel}=2.9$ (RN) (see Fig.~\ref{fig4}).
  Molecules are aligned with $\widehat{\bm{n}}$ parallel to the x axis,
  whereas flat surfaces on top and bottom in dark represent the solid
  substrates.}\label{fig5}
\end{figure}

At $T=4.0$ there is a clear sign of a RN phase as $S$ increases and $\Lambda$
reaches a maximum value and then drops to small values for
$P_{\parallel}\simeq2.3$; thus, a sequence of phases I-N-S$_\mathrm{A}$-RN is
found at $T=4.0$ as $P$ increases. At $T=6.0$, $\Lambda$ never rises above the
residual value of about $0.1$. Hence, the S$_\mathrm{A}$ phase does not form
at this $T$.  Nevertheless, we notice a small, step-like increase at
$P_{\parallel}\simeq2.2$ in the plot of $\Lambda$. At this and all larger
pressures considered the nematic order is high and increases even further
reflected by a monotonic increase of $S$ toward its limiting value $1.0$. For
$P_{\parallel}\gtrsim2.2$ the confined fluid exhibits structural features of
the RN phase (see \cite{mazza11}).  Three characteristic snapshots of the
system are shown in Fig.~\ref{fig5}. The increasing orientational order of
the system as $P_{\parallel}$ increases is apparent.
Incidentally, we note that the seemingly value $S=1$ in the RN phase is a
finite-size effect, as was demonstrated by Eppenga and
Frenkel~\cite{eppenga84}.


\begin{figure}
\begin{center}
\includegraphics[width=0.6\linewidth]{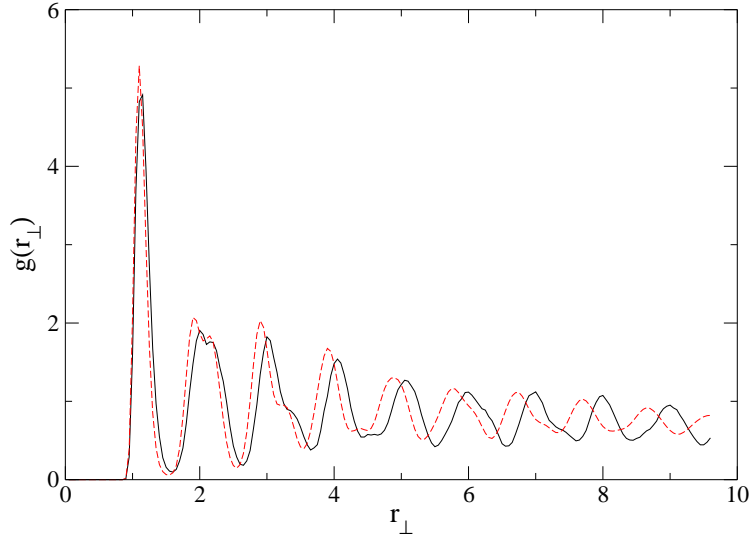}
\end{center}
\caption{(Color online) Perpendicular radial distribution function for a bulk
  (solid line) and confined (dashed line) simulation. Both states are in the
  RN phase at $T=4.0$.}\label{g_perp}
\end{figure}

A question naturally arises: What is the relation between the RN phase found
in this simulated confined fluid and the bulk RNs? Our simulations show that
the phase diagram of the confined system is qualitative similar to phase
diagram of the bulk fluid. The only change induced by the confining surfaces
is to lower the pressure of the phase transitions with respect to the bulk
system. This shift in the phase diagram is a well-documented confinement
effect. Here, we consider spatial correlations between the centers of mass of
molecules in the direction $\bm{r}_\perp$ perpendicular to the orientation of
a reference molecule, where
$\bm{r}_\perp\equiv\bm{r}-(\widehat{\bm{u}}\cdot\bm{r})\widehat{\bm{u}}$. The
perpendicular radial distribution function $g(r_\perp)$ is the probability to
find a molecule at distance $r_\perp=|\bm{r}_\perp|$ from a reference
molecule.  In Fig.~\ref{g_perp} we compare $g(r_\perp)$ for a bulk and a
confined system in the RN phase. It is apparent that the maxima of
$g(r_\perp)$ for the confined system are shifted to lower values of $r_\perp$
with respect to the bulk case. This shift reflects the substrate-assisted
efficient packing of molecules and therefore explains the shift of the
S$_\mathrm{A}$-RN transition to lower $P_\parallel$ than in the bulk. However,
the structure of the confined and bulk RN phases are very similar.

Next, we address the important question of dynamics in the RN phase. To this
end, we considered a {\em specialized} mean square displacement (MSD) to
calculate displacements of molecules in the direction of their long axes.
Specifically, we define $r_i^{\parallel}\equiv\widehat{\bm{u}}_i\cdot\bm{r}_i$
and the associated MSD in the direction of the molecular long axes
\begin{equation}\label{eq:msd}
\left\langle
\Delta r_{\parallel}^2\left(\tau\right)
\right\rangle_{t}
\equiv
\frac{1}{N}
\left\langle
\sum\limits_{i=1}^{N}
\left[
r_i^{\parallel}\left(t+\tau\right)-r_i^{\parallel}\left(t\right)
\right]^2
\right\rangle_{t}\,.
\end{equation}
From Eq.~\ref{eq:msd} a diffusion coefficient can be extracted using an
Einstein relation, namely
\begin{equation}
D_{\parallel}=
\lim\limits_{\tau\to\infty}
\frac{1}{2\tau}
\left\langle
\Delta r_{\parallel}^2\left(\tau\right)
\right\rangle_{t}\,.
\end{equation}

\begin{figure}
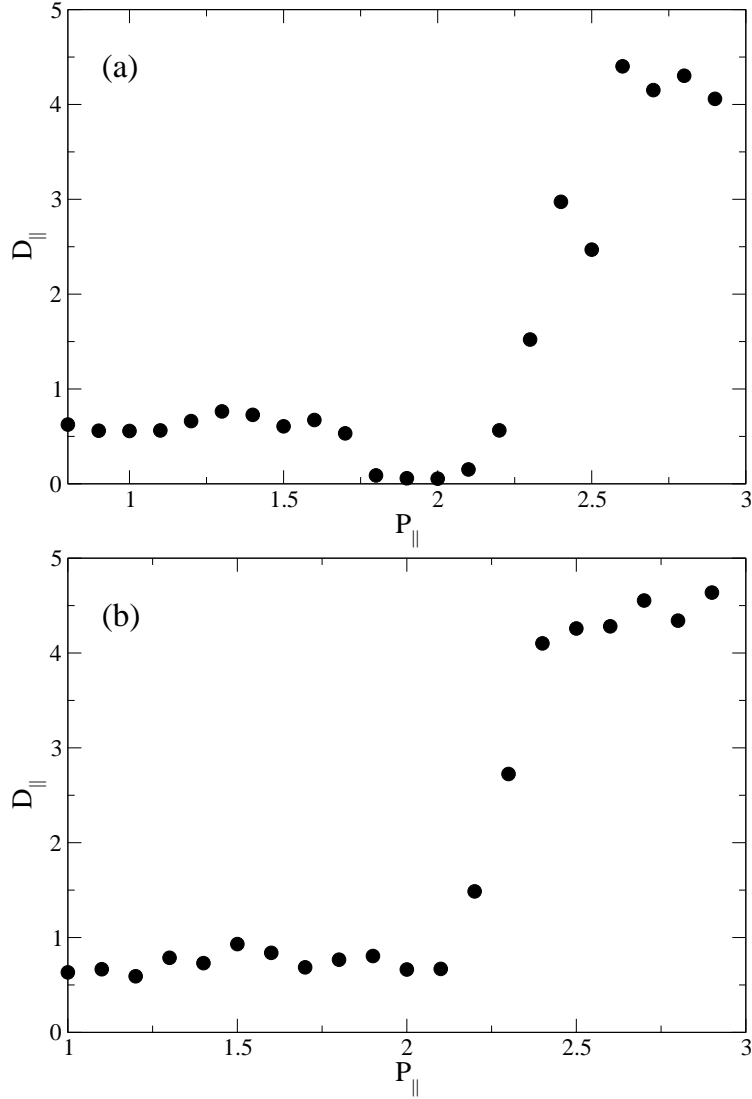

\begin{center}
\includegraphics[width=0.6\linewidth]{D_par-CompAFxy-T4.0.eps}
\includegraphics[width=0.6\linewidth]{D_par-CompAFxy-T6.0.eps}
\end{center}
\caption{\small (Color online) Parallel self-diffusion coefficient
  $D_{\parallel}$ as a function of transverse pressure $P_{\parallel}$.  (a)
  Calculations at $T=4.0$; (b) Calculations at $T=6.0$.}\label{fig12}
\end{figure}

Figure~\ref{fig12} shows a dramatic increase of $D_{\parallel}$ as one enters
the RN phase whereas lower-pressure (I, N, or S$_\mathrm{A}$) phases exhibit
rather small self-diffusivity. In particular, the S$_\mathrm{A}$ phase is
characterized by nearly vanishing self-diffusion constants which can be
rationalized as above where we argued that the relatively compact layered
structure makes it difficult for molecules to diffuse out of their original
layer and penetrate into a neighboring one. The dramatic increase in mass
transport in the direction of $\widehat{\bm{n}}$ in combination with nearly
perfect nematic order prompted us to refer to liquid crystals in the RN phase
as ``supernematics'' \cite{mazza10}.

There are, in fact, experimental observations that are consistent with our
simulations.  For example, by extracting longitudinal relaxation rates from
data of Miyajima \emph{et al.}~\cite{miyajima84}, one finds that in the RN
phase these relaxation rates are considerably lower than those characteristic
of the N phase. At the same $T$, relaxation rates can be converted into
correlation times that are significantly shorter in the RN as opposed to the N
phase.  However, this remains speculative, until experiments prove that
translation diffusion is the main relaxation mechanism at this $T$.  Stronger
evidence is found in the conductivity experiments performed by Ratna
\cite{ratna81} (see Sec.~\ref{exp}).

One may ask: Why is the diffusivity so large in the direction of the long
molecular axis? As interesting analogy \cite{mazza11} can be drawn with the
case of diffusion in systems such as zeolites or single-wall carbon nanotubes.
When the diameter of a diffusing particle matches the size of the confining
nanochannel then the diffusivity increases markedly. This effect is known as
``levitation''~\cite{derouane,yash}. In our case, the high degree of
orientational order in the RN phase causes a molecule to be trapped within the
``cage'' formed by its first-neighbors which, because of the prolate geometry,
is effectively a narrow channel. Furthermore, the attractive intermolecular
forces are overcome because the fluid is in a dense state (either low $T$ or
high pressure). Thus, within this channel a molecule fits the condition of the
levitation effect.

\section{Conclusions}\label{concl}

In this work, we have reviewed experimental and theoretical investigations of
structural and dynamical properties of RN phases. It is surprising that after
almost $40$ years of research many details of RN phases are still escaping a
comprehensive understanding.  The main difficulty is the variety of substances
exhibiting RN's. Pure substances and mixtures alike have been found with a RN
in their phase diagrams. Polar and nonpolar molecules exhibit reentrancy
alike.  It is clear to the present authors that more experiments are necessary
to settle some important, yet still open questions. For example: (i) How does
the nematic order change upon crossing the S$_\mathrm{A}$-RN transition?  (ii)
Is frustration really the mechanism underlying the reentrant behavior in LC's
with a bilayer S phase?  (iii) Experimentally, what is the mechanism driving
reentrancy in LC's that do not form dimers? (iv) Do RN phases of
single-component, nonpolar molecules really exhibit the huge self-diffusivity
in the direction of $\widehat{\bm{n}}$ predicted by our simulations
\cite{mazza10, mazza11}? 

Regarding the dynamical behavior, it is at present safe to assume that though
many LC systems have a RN phase, their dynamics need not be alike in any
respect.  The reason is that different microscopic mechanisms may be
responsible for reentrancy in different LC systems depending on the details of
the intermolecular interaction potential. At the time of writing some
experimental evidence exists \cite{ratna81} suggesting that last question may
indeed have a positive answer. Moreover, our findings \cite{mazza10, mazza11}
seem consistent with an extrapolation of longitudinal relaxation rates of NMR
experiments \cite{miyajima84}.

\section*{Acknowledgments}
We thank M. Greschek (Technische Universit\"at Berlin), J. K\"arger
(Universit\"at Leipzig) and R. Valiullin (Universit\"at Leipzig) for helpful
discussions.  Financial support by the German Research Foundation (DFG) within
the framework of the ``International Graduate Research Training Group'' 1524
is gratefully acknowledged.

\bibliographystyle{mdpi}
\makeatletter
\renewcommand\@biblabel[1]{#1. }
\makeatother


\begin{thebibliography}{-------}
\bibitem[not()]{note1}
A spherocilinder is defined as a cylinder of length $L$ and diameter $\sigma$
  capped at both ends with hemispheres of diameter $\sigma$. Thus, the total
  length of the spherocilinder is $L+\sigma$.
\end{thebibliography}

\end{document}